\documentclass[final,1p,times]{elsarticle} 

\usepackage{graphicx}
\usepackage{amssymb} 
\usepackage{amsthm} 
\usepackage{lineno}

\journal{Nuclear Physics A} 

\begin{document}

\begin{frontmatter} 

\title{D meson nuclear modification factors in Pb--Pb collisions\\ at $\sqrt{s_{\rm NN}}~=~2.76~{\rm TeV}$ with the ALICE detector}

\author{Alessandro Grelli (for the ALICE\fnref{col1}  Collaboration)}
\fntext[col1] {A list of members of the ALICE Collaboration and acknowledgements can be found at the end of this issue.}
\address{ERC- Research Group QGP, Utrecht University, Princeton plein 5, 3584 CC Utrecht, The Netherlands}


\begin{abstract} 
The measurement of D meson production provides key tests for parton energy-loss models, which predict  that charm quarks should experience less in-medium energy loss than light quarks and gluons.
The ALICE experiment has measured the production of prompt $\rm D^0$, $\rm D^+$ and $\rm D^{*+}$ mesons in pp and Pb--Pb collisions at the LHC at $\sqrt{s}$~=~7 and 2.76 $\rm TeV$ and at $\sqrt{s_{\rm NN}}$~=~2.76~{\rm TeV}, respectively, via the exclusive reconstruction of their hadronic decay. The $p_{\rm T}$-differential production yields in the range 2 $< p_{\rm T}<$ 16 ${\rm GeV}/c$ at central rapidity, $|y| < 0.5$, were used to calculate the nuclear modification factor. A suppression of a factor 3 to 4 for transverse momenta larger than 5 ${\rm GeV}/c$ in the 20$\%$ most central collisions was observed. Preliminary results in an extended $p_{\rm T}$-range, using the data sample collected during the 2011 Pb--Pb run, together with the first measurement of $\rm D^+_{\rm s}$ nuclear modification factor will be shown.
\end{abstract} 

\end{frontmatter} 


\section{Introduction}
\label{intro}
Under conditions of high-energy density and temperature, produced in ultra-relativistic heavy-ion collisions, lattice QCD calculations predict the transition from ordinary nuclear matter to a deconfined state of quarks and gluons called Quark Gluon Plasma (QGP)~\cite{w}. Heavy-flavour quarks, differently from light quarks and gluons, are only produced at the early stage of the collision in high-virtuality scattering processes. They traverse the medium and are expected to be sensitive to its density through the mechanism of in-medium partonic energy loss. Heavy flavours should lose less energy than light-quarks and  gluons as a consequence of a mass-dependent restriction in the phase space into which gluon radiation can occur~\cite{eloss3}. Moreover, if in-medium hadronization is the dominant mechanism of charm-hadron formation then the strange charmed hadrons, like $D^+_{\rm s}$, should be largely enhanced~\cite{Ds}.
The nuclear modification factor ($R_{\rm AA}$) of D mesons, obtained by comparing their production in proton-proton and heavy ion collisions, allows to probe the aforementioned properties. The $R_{\rm AA}$ is defined as: $R^{{\rm D}}_{{\rm AA}}(p_{\rm T}) =\frac{1}{\langle \rm T_{{\rm AA}}\rangle} \frac{{\rm d}N^{\rm D}_{{\rm AA}}/ \rm d p_{\rm T}}{\rm d\sigma^{\rm D}_{{\rm pp}}/\rm d p_{\rm T}}$ where $N^{\rm D}_{{\rm AA}}$ is the yield in A--A collisions, $\langle \rm T_{{\rm AA}}\rangle$, in a given centrality class, is the average nuclear overlap function calculated via Glauber model and $\sigma^{\rm D}_{\rm pp}$ is the production cross section of D mesons in pp collisions. In absence of medium effects $R^{\rm D}_{\rm AA}$  is expected to be 1. Using the data collected in 2010 we published the first measurement of the suppression of high transverse momentum D mesons in central Pb--Pb collisions at $\sqrt{s_{\rm NN}}~=~2.76~{\rm TeV}$~\cite{PbPb_2010}. We observed a suppression of a factor 3 to 4 for transverse momenta larger than 5 ${\rm GeV}/c$ in the 20$\%$ most central collisions. In this contribution we present the nuclear modification factor of D$^0$, D$^+$, D$^{*+}$ and ${\rm D}^+_{\rm s}$ as function of transverse momentum in the centrality range 0-7.5$\%$ as extracted from the data collected in 2011. A comparison with model predictions is made.   

\section{D meson reconstruction with ALICE}
\label{data}

The production of ${\rm D}^0$, ${\rm D}^+$, ${\rm D}^{*+}$~\cite{PbPb_2010,Dpp7paper} and ${\rm D}^+_{\rm s}$~\cite{Dss,Dss1} was measured in pp and Pb--Pb collisions at central rapidity ($|y|<0.5$) via the exclusive reconstruction of the decays ${\rm D}^0\rightarrow {\rm K}^-\pi^+ $, ${\rm D}^+\rightarrow {\rm K}^-\pi^+\pi^+$, ${\rm D}^{*+}\rightarrow {\rm D}^0\pi^+\rightarrow {\rm K}^-\pi^+\pi^+$ and ${\rm D}^+_{\rm s}\rightarrow \phi \pi^+\rightarrow {\rm K}^-{\rm K}^+\pi^+$. The analysis strategy for the extraction of the signal out of a large combinatorial background is based on the reconstruction of the D meson decay topology (${\rm D}^0$ in case of the ${\rm D}^{*+}$). In order to resolve the vertex of the decay the Inner Tracking System (ITS)~\cite{aliceJINST} provides the required resolution of a few tens of microns on the track position at the primary vertex. D meson candidates are defined from displaced tracks and selected by means of topological cuts. To further suppress the combinatorial background, particle identification (PID) on the D meson decay tracks is employed. PID is performed using the information on specific energy deposit in the Time Projection Chamber (TPC) and on the time of flight measured by the Time of Flight (TOF) detector. It allows a reliable identification of kaons and pions in a wide momentum range up to 2 GeV/${\rm c}$. The signal yield is extracted by fitting the invariant mass distribution using a Gaussian function for the signal peak. The background is fitted using an exponential shape in the case of ${\rm D}^0$, ${\rm D}^+$ and ${\rm D}^+_s$ while in the case of the ${\rm D}^{*+}$ a threshold function convoluted with and exponential is chosen. The correction for efficiency and acceptance is performed using Monte Carlo simulations based on Pythia Perugia-0 tuning and HIJING event generator. In order to extract the prompt D fraction, the contribution of D mesons from B decays was evaluated relying on FONLL~\cite{FONLL} calculations, which describe well B hadron production at Tevatron~\cite{fff} and LHC~\cite{ddd,ddd1}. A data driven method, based on the different shape of the impact parameter to the primary vertex for primary and secondary D mesons is under study. The analyses presented in this paper are based on the Pb--Pb data sample at centre-of-mass energy $\sqrt{s_{\rm NN}}~=~2.76$ TeV collected in November 2011. The events were triggered with centrality-based selection using information from Silicon Pixel Detector ($|\eta|<2$) and the VZERO~\cite{aliceJINST}  scintillator hodoscopes  ($2.8<\eta<5.1$ and $-3.7<\eta<-1.7$). Only events with a vertex found within 10~cm from the centre of the detector along the beam line were used, for a total of $16\times 10^6$ collisions in the 0-7.5$\%$ centrality class.
 
 \section{D meson pp reference}
 
The D-meson production cross section in pp collisions at $\sqrt{s}=2.76$ TeV, needed for the $R_{\rm AA}$ calculation, is based on a scaling of the measured D meson production cross section at $\sqrt{s}=7$ $\rm TeV$ using the ratio of FONLL predictions at $\sqrt{s} =$ 2.76 and 7 $\rm TeV$. The procedure is validated with data by comparing the scaled cross section with the one measured on a pp sample with limited statistics at $\sqrt{s}$ = 2.76 $\rm TeV$. A detailed explanation of the procedure can be found in~\cite{276_2010}.  At high $p_{\rm T}$, $24<p_{\rm T}<36~{\rm GeV}/c$ for  ${\rm D}^+$, ${\rm D}^{*+}$ and $16< p_{\rm T}<24~{\rm GeV}/c$ for  ${\rm D}^0$  a measured reference is not available. Therefore the cross section was extrapolated on the basis of the FONLL/data ratio observed at lower transverse momentum. In $1<p_{\rm T}<2~{\rm GeV}/c$, where the total uncertainties of the 7 $\rm TeV$ scaling and of the 2.76 $\rm TeV$ measurement are comparable, the reference was calculated as the weighted average of the 7 TeV scaled and of the 2.76 TeV measurement using the relative uncorrelated uncertainties as weights.

\section{D meson nuclear modification factors}

\begin{figure}[b!]
\centering
 \begin{minipage}{5.95cm}\noindent
\includegraphics[width=0.85\textwidth]{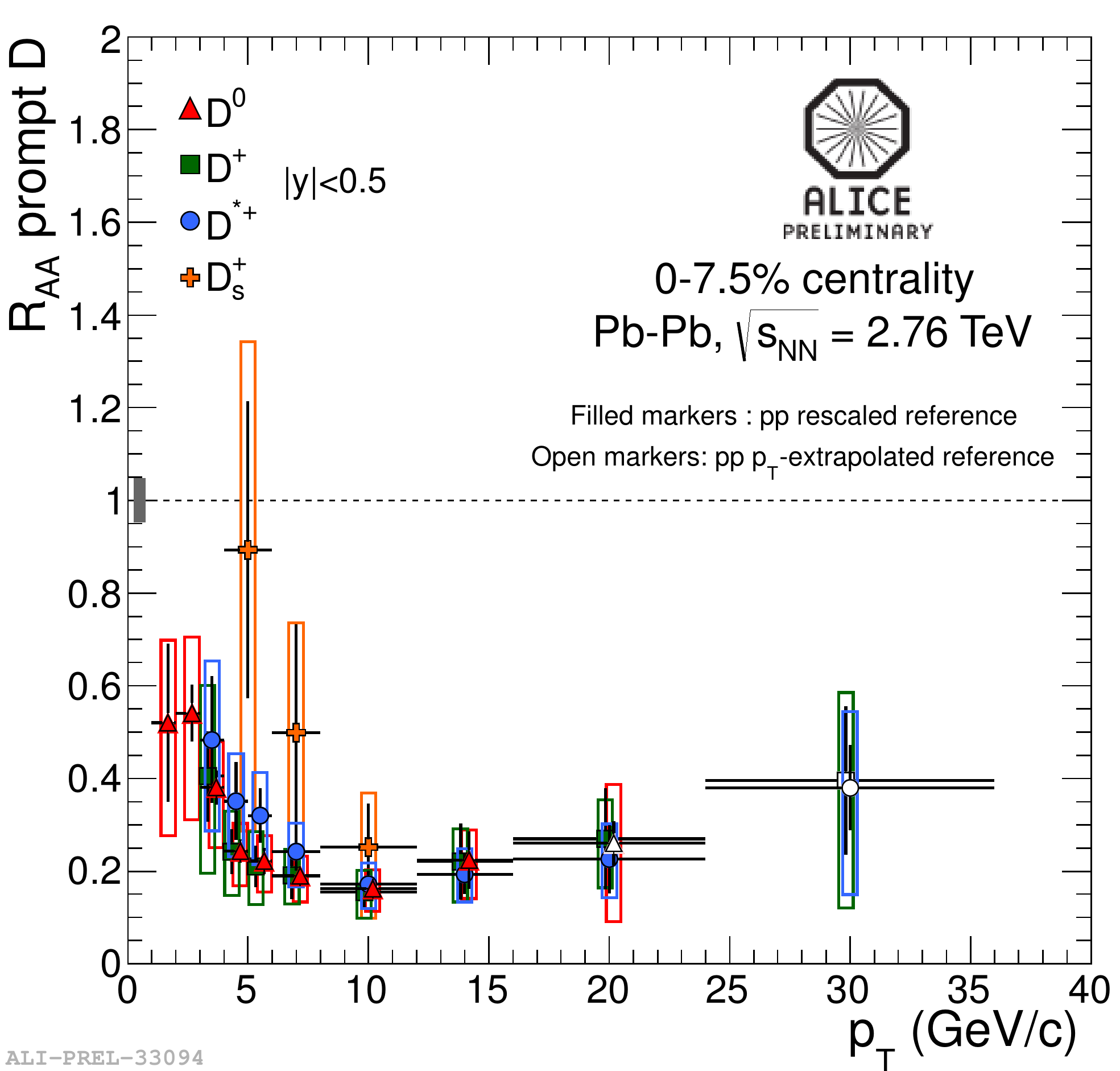}
 \end{minipage}%
  \begin{minipage}{5.96cm}\noindent
\includegraphics[width=1.04\textwidth]{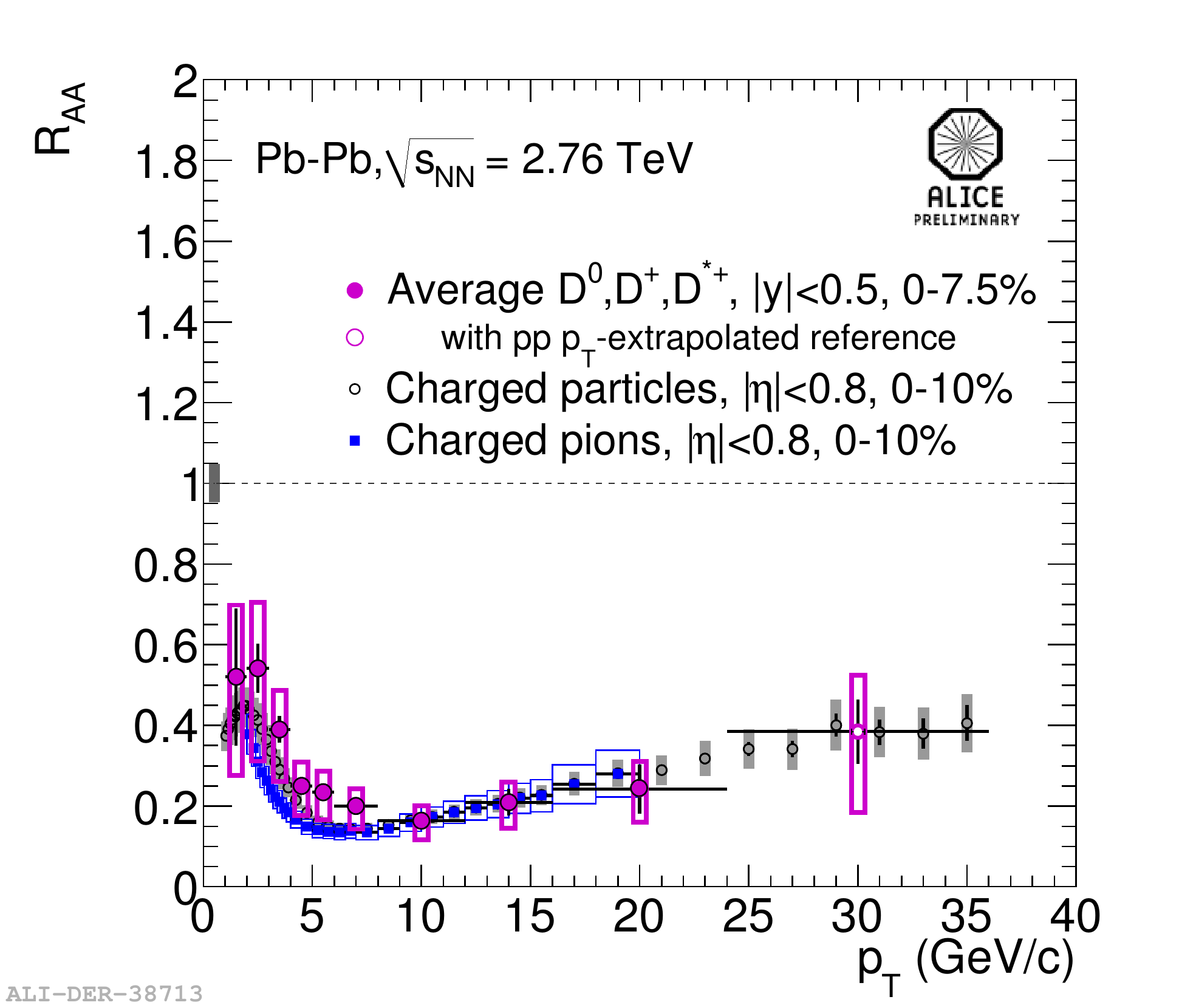}
\end{minipage}%
\caption{Left panel: ${\rm D}^0$, ${\rm D}^+$, ${\rm D}^{*+}$ and ${\rm D}^+_{\rm s}$ meson nuclear modification factor in the centrality class 0-7.5$\%$ in the $p_{\rm T}$ range $1<p_{\rm T}<36~{\rm GeV}/c$ ($4<p_{\rm T}<12~{\rm GeV}/c$ in the case of D$^+_{\rm s}$). Right panel: Comparison of the average ${\rm D}^0$, ${\rm D}^+$ and ${\rm D}^{*+}$ $R_{\rm AA}$ with charged pions and charged hadrons.} 
\label{fig1}
\end{figure}
\begin{figure}[t!]
\centering
\includegraphics[width=0.89\textwidth]{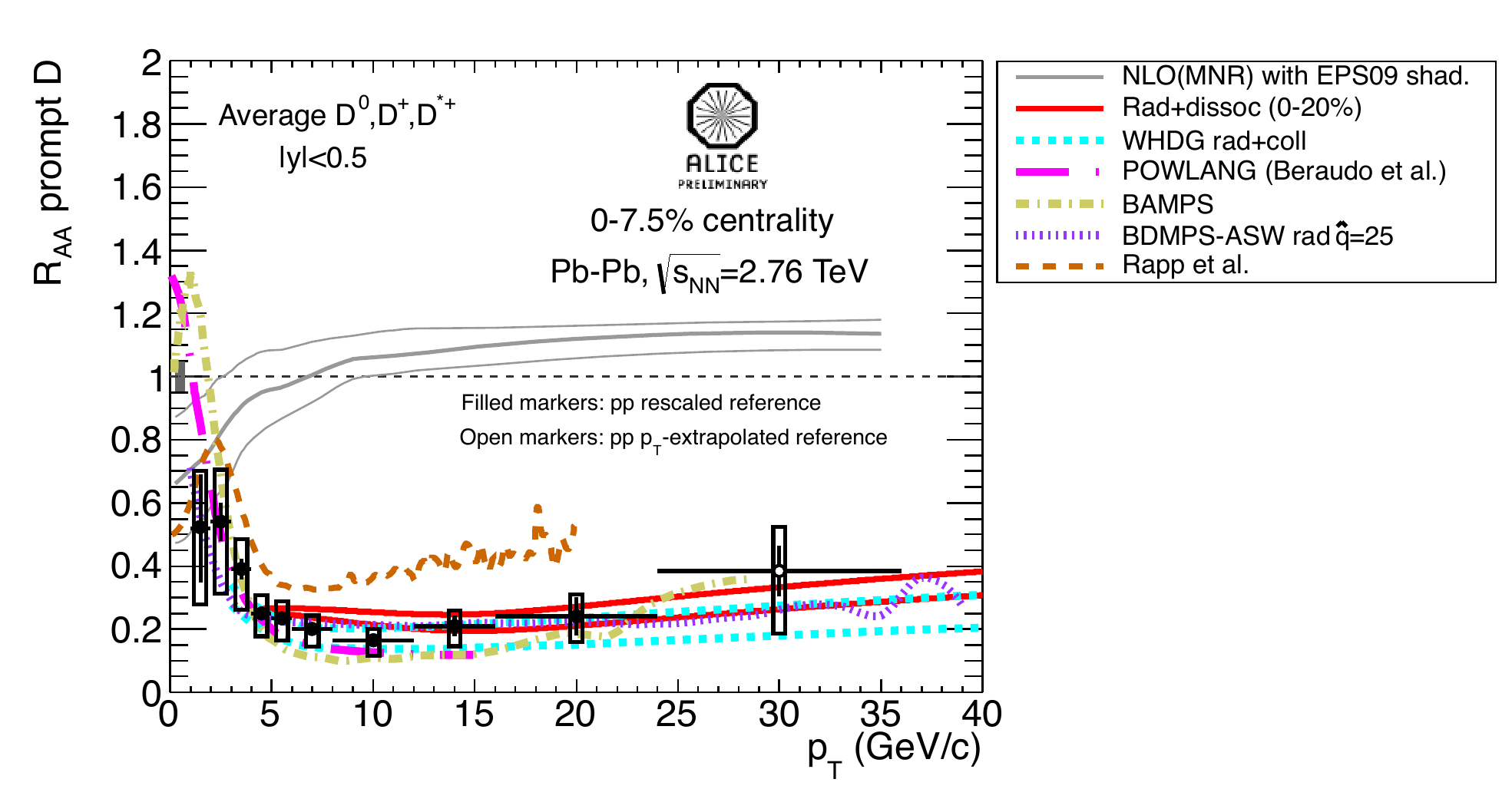}
\caption{Average $R_{\rm AA}$ of D$^0$, D$^+$ and D$^{*+}$ compared with model calculations} 
\label{dmes}
\end{figure}

The nuclear modification factor of D$^0$, D$^+$, D$^{*+}$ and D$^+_{\rm s}$ mesons measured with the data taken in 2011 in the centrality class 0-7.5$\%$ is shown on the left panel of Fig.~\ref{fig1}.~The $R^{\rm D}_{{\rm AA}}$ of the four D mesons agrees within uncertainties and shows a clear suppression of about a factor 5 for $p_{\rm T}\sim~10~{\rm GeV}/c$. The estimated total systematic uncertainty, accounting for the uncertainties on signal extraction procedure, PID selection strategy, track reconstruction efficiency, cut stability and hypothesis on B mesons $R_{\rm AA}$ (accounting for all the potential nuclear and medium effects affecting B production) varies from $\sim$70$\%$ to 12$\%$ depending on $p_{\rm T}$ and D meson species. The $R_{\rm AA}$ of D$^0$, D$^+$, and D$^{*+}$ were averaged.~The contributions of the three D mesons to the average were wighted by their statistical uncertainties. The systematic errors were calculated by propagating the uncertainties through the weighted average, where the contributions from the tracking efficiency, B feed-down correction and scaling of the pp reference were taken as fully correlated among the three D meson species.~On the right panel of Fig.~\ref{fig1} we compare our result with charged pions and charged hadrons  $R_{{\rm AA}}$ as measured by ALICE with 2011 data sample in the centrality class 0-10$\%$. The three measurements are compatible within uncertainties, however a hint of a lower suppression of the D mesons is seen at low $p_{\rm T}$. The data points show a hint for $R^{\rm D}_{{\rm AA}} < 1$ at low $p_{\rm T}$ while at high $p_{\rm T}$ the data are consistent with both a flat and a rising shape.~Fig.~\ref{dmes}  shows the average $R^{\rm D}_{{\rm AA}}$ compared with models.~We found that radiative energy loss supplemented with in-medium D meson dissociation~\cite{radd} and radiative plus collisional energy loss in the WHDG~\cite{WHDG} implementations, together with BDMPS-ASW~\cite{BDMPS} and  POWLANG~\cite{rad} describe the data reasonably well.~The heavy quark transport model with in-medium resonance scattering and coalescence~\cite{rapp} tends to underestimate the suppression observed in the data. We verified, using pQCD calculations based on the MNR~\cite{MNR} code and nuclear-modified PDF from the EPS09~\cite{EPS} parametrization, that nuclear shadowing has a relatively small effect for $p_{\rm T}\sim 10~{\rm GeV}/c$.~This indicates that the suppression, of about factor 5 at $p_{\rm T} \sim 10~{\rm GeV}/c$ indicate a large energy loss of charm quarks in the hot and dense QCD medium formed in heavy-ion collisions at the LHC.

\section{Summary}

We have presented the first measurement of the D$^0$, D$^+$, D$^{*+}$ nuclear modification factor in the centrality class 0-7.5$\%$ using the 2011 Pb--Pb data. The $p_{\rm T}$ range of the measurement is extended with respect to the 2010 result~\cite{PbPb_2010} and now covers the transverse momentum range $1< p_{\rm T}< 36~{\rm GeV}/c$. The p--Pb run scheduled for the beginning of 2013 will allow for the study of the initial state effect at low $p_{\rm T}$. Finally we presented the first $R_{\rm AA}$ of the ${\rm D}^+_{\rm s}$ meson, in the $p_{\rm T}$ range $4<p_{\rm T}<12~{\rm GeV}/c$,  together with its comparison with the other D mesons. The result of the comparison, not conclusive due to the large statistical and systematic uncertainties, is intriguing in view of future LHC runs.

\bibliographystyle{model2-names}
\bibliography{bib}

\begin{thebibliography}{00}
 
 
 \bibitem{w}
D.~J.~Gross and F.~Wilczek, Phys. Rev. Lett. 30, 1343 (1973), H.~D.~Politzer, Phys. Rev. Lett. 30, 1346 (1973)

\bibitem{eloss3}
A.~Dainese,~N.~Arnesto,~M.~Cacciari,~C.~A.~Salgado and U.~A.~Wiedemann, AIP Conf. Proc., 842 (2006) pp. 74-76

\bibitem{Ds}
I.~Kuznetsova and J.~Rafelski, Eur.Phys.J. C51 (2007) 113-133. M.~He, R. J.~Fries and R. Rapp, arXiv:1204.4442
 
 \bibitem{PbPb_2010}
B.~Abelev et al. [ALICE Collaboration] JHEP {\bf 09} (2012) 112 

\bibitem{Dpp7paper}
B.~I.~Abelev {\it et al.} [ALICE Collaboration], JHEP {\bf 01} (2012) 128

\bibitem{Dss}
B.~I.~Abelev {\it et al.} [ALICE Collaboration], CERN-PH-EP-2012-227, arXiv:1208.1948

\bibitem{Dss1}
G.~Innocenti  [ALICE Collaboration], proceeding, this conference.

\bibitem{aliceJINST}
K.~Aamodt {\it et al.} [ALICE Collollaboration], JINST {\bf 3} (2008) S08002

\bibitem{FONLL} 
 M.~Cacciari, M.~Greco and P.~Nason, JHEP 9805 (1998) 007;
M.~Cacciari {\it et al.} JHEP 0103 (2001) 006
 
 \bibitem{fff} 
 M.~Cacciari {\it et al.}, JHEP 0407 (2004) 033
 
  \bibitem{ddd} 
R.~Aaij {\it et al.} [LHCb Coll.], Phys. Lett. B694 (2010) 209

 \bibitem{ddd1} 
 V.~Khachatryan {\it et al.} [CMS Collaboration], Eur. Phys. J. C71 (2011) 1575
 
\bibitem{276_2010}
B.~Abelev {\it et al.} [ALICE Collaboration],  JHEP {\bf 07} (2012) 191 

\bibitem{radd}
R.~Sharma {\it et al.}, Phys. Rev. C80 (2009) 054902. Y. He {\it et al.}, arXiv:
1105.2566 [hep-ph] (2011).

\bibitem{WHDG}
W.~A.~Horowitz and M.~Gyulassy, J. Phys. G38 (2011) 124114

\bibitem{BDMPS}
N.~Armesto, A.~Dainese, C.~A.~Salgado and U.~A.~Wiedemann, Phys. Rev. D71 (2005) 054027

\bibitem{rad}
A.~Beraudo {\it et al.}, J. Phys. G G38 124144

\bibitem{rapp}
M. He {\it et al.}, arXiv:1106.6006.

\bibitem{MNR}
M. Mangano, P. Nason and G. Ridolfi, Nucl. Phys. B373 (1992) 295.

\bibitem{EPS}
K. J. Eskola, H. Paukkunen and C. A. Salgado, JHEP 04 (2009) 065.


 \end{thebibliography}

\end{document}